\documentclass[twocolumn,showpacs,preprintnumbers,amsmath,amssymb]{revtex4}

\usepackage{graphicx}

\usepackage{bm}
\usepackage{booktabs}
\usepackage{dcolumn}

\usepackage{slashed}

\newcommand{\etal}{{\it et al}}

\begin{document}

\title{$D_s(0^\pm)$ Mesons spectroscopy in Gaussian Sum Rules
\footnote{Supported by the National Natural Science Foundation of
China under Grant No. 10775105, BEPC National Laboratory Project
R\&D and BES Collaboration Research Foundation.}}

\author{WEN Shuiguo （闻水国）}

\author{LIU Jueping （刘觉平）
\footnote{To whom correspondence should be addressed. Email:
jpliu@whu.edu.cn}}

\affiliation{%
College of Physics and Technology, Wuhan University, Wuhan
430072}%

\date{\today}

\begin{abstract}

The masses of the $D_s(0^\pm)$ mesons are investigated from a
view-point of ordinary light-heavy system in the framework of the
Gaussian sum rules, which are worked out by means of the Laplacian
transformation to the usual Borel sum rules. Using the standard
input of QCD non-perturbative parameters, the corresponding mass
spectra and couplings of the currents to the $D_s(0^\pm)$ mesons are
obtained. Our results are $m_{D_s(0^-)}=1.968\pm0.016\pm0.003$ GeV
and $m_{D_s(0^+)}=2.320\pm0.014\pm0.003$ GeV, which are in
accordance well with the experimental data, 1.969 GeV and 2.317 GeV.

\end{abstract}


\maketitle

Recently, the experimental observations of $D$ meson and the
corresponding theoretical manipulations attract much attention in
the research of particle
physics\cite{Drutskoy:2008,Drutskoy:2005,Mikami:2004,Aubert:2004hb,Abe:2004hk,Vaandering:2004he,
Krokovny:2003mh, Aubert:2003ha,Besson:2003hx,Eidelman:2004mh}. Many
Quantum Chromodynamics (QCD) calculations about $D$ mesons are in
good agreement with the experimental data within the theoretical
uncertainty except for the unexpected low mass of  $D_{0^+}(2317)$
\cite{Hayashigaki:2004hj, Pierro:2001hm,
Narison:2001mh,Narison:2005prb}. In Ref. \cite{YBDai:2003}, the
masses of the excited $(0^+,1^+)$ and $(1^+,2^+)$ doublets for the
$c\bar{s}$ system are calculated to the $1/m_c$ order in the sum
rules based on heavy quark effective theory(HQET) , and the
$D_{s}(2317)$ and $D_{s}(2460)$ are identified as the $(0^+,1^+)$
doublet so that the mass splitting in this doublet is well
reproduced. We note that most of these theoretical calculations are
based on QCD Borel sum rule (BSR) which emphasize the contributions
of lowest resonance state, and have shown the power in the
investigation of the non-perturbative properties of hadron bound
states. However, excited states and continuous spectrum can produce
background interference. Generally, this background interference
increases with the mass of the hadron state considered, and this
restrict the application range of BSRs. It is noticed that the QCD
Gaussian sum rule (GSR) developed later emphasizes only the
contribution of the hadron state considered as seen from the
appearance of the Gaussian distribution function, and has more clean
background in comparison with the BRS. From this reason, the GSR
may, in principle, work better than the BSR. At least, both sum
rules should give almost the same results because they are derived
from the same underlying dynamical theory.

On the other hand, whether the $D_s$ meson's spectroscopy observed
by experiments can be derived out from the elementary theory of the
strong interactions is a key point for testing the QCD
non-perturbative mechanism, and serves also to be the starting point
of calculating the important physical processes, such as the $D_s$
meson decays and etc, from the first principle. However, the results
obtained by using the BSRs, $m_{D_s(0^+)}=2.48\pm0.03$ GeV and
$m_{D_s(0^-)}=1.94\pm 0.03$ GeV \cite{Hayashigaki:2004hj}, are
obvious inconsistent with the experiment data,
$m_{D_s(0^+)}=2.317\pm0.0013$ GeV\cite{Aubert:2003ha} and
$m_{D_s(0^-)}=1.969$ GeV. For checking the correctness of QCD and
exploring the non-perturbative mechanism, it's necessary to do a
further study.

In order to make a cross check between various types of QCD sum
rules, namely the BSRs\cite{Hayashigaki:2004hj} and the HQET sum
rules\cite{YBDai:2003}, we calculate the masses $m_{D_s}$ of $D_s$
mesons and the couplings $f_{D_s}$ of the corresponding currents to
these mesons by using GSRs.

Consider the gauge-invariant and Lorentz-covariant two-point
function of the current $J_i(x)$ corresponding to a resonance with
the quantum numbers $i=0^-,0^+$ \cite{Reinders:1985im,
Shifman:1979ik}
\begin{equation}
 \Pi_i(q^2)=i\int
d^4xe^{iq\cdot x}\langle0|TJ_i(x)J_i^\dagger(0)|0\rangle,
\end{equation}
where $|0\rangle$ is nonperturbative QCD vacuum, and $J_i(x)$ are
the pseudoscalar and scalar currents
\begin{eqnarray}
J_{0^-}(x)&=&i\bar{q}(x)\gamma_5c(x),\\
J_{0^+}(x)&=&\bar{q}(x)c(x),
\end{eqnarray}
with $q(x)$ and $c(x)$ being the strange-quark and charm-quark
fields at the point \textit{x}, respectively. For the invariant
functions $\Pi(q^2)=\Pi_i(q^2)(i=P,S)$, we have a dispersion
relations without any subtraction
\begin{equation}
\Pi(q^2)=\frac1{\pi}\int ds\frac{\textrm{Im}
\Pi(s)}{s-q^2+i\epsilon}.\label{eq:dispe}
\end{equation}

Via this dispersion relation, the QCD sum rule (QSR) was constructed
by equating the contribution of operator product expansion (OPE) and
the phenomenological (PH) one related to the spectral function. The
former is described as the product of Wilson coefficients and
nonperturbative QCD vacuum condensates or quark masses, while the
latter is parameterized by hadronic quantities such as resonance
masses, couplings and the continuum threshold, $etc$. Thus, the QSR
can be represented in a simple form,
\begin{equation}
\int^\infty_{m^2_c}dsW(s)\frac1{\pi}\left[\textrm{Im}\Pi^{\textrm{PH}}(s)-\textrm{Im}\Pi^{\textrm{OPE}}(s)\right]=0\label{eq:QSR},
\end{equation}
where $W(s)$ is an arbitrary weight function being analytic except
for the positive real axis starting from the lower mass squared,
$m^2_c$. According to the most successful application of sum rules
to mesons and baryons\cite{Reinders:1985im}, the phenomenological
spectral function is assumed to be saturated by one narrow-width
resonance and a continuum
\begin{equation}
\frac{1}{\pi}\textrm{Im}\Pi^{\textrm{PH}}(s)=F\delta(s-m^2_R)+\frac{1}{\pi}\textrm{Im}\Pi^{\textrm{OPE}}(s)\theta(s-s_0),\label{eq:Spect}
\end{equation}
where $s_0$ is the QCD continuum threshold, the pole residue is of
the form $F=f^2_Rm^{2k}_R$ with $f_R$ being the couplings of the
lowest resonances with respective parities to the hadronic currents
and $m_R$ a pole mass. The power \textit{k} of $m^2_R$ in the pole
residue is taken to match the maximum power of \textit{s} in the
asymptotic \textit{s}-behavior of the spectral function. For
$s>s_0$, the hadronic continuum reduces to the same form with that
obtained by an analytic continuation of the
OPE\cite{Hayashigaki:2004hj}, i.e. the perturbative terms, based on
a hypothesis of the quark-hadron duality.

Implementing the Borel transformation to the QSRs and performing the
OPE at dimension $d\leq 6$ operators, the relations of the BSRs of
the lowest $0^{\pm}$ $c\bar{s}$-mesons are obtained to be
\cite{Hayashigaki:2004hj}
\begin{eqnarray}
&&f_{0^\pm}^2 m_{0^\pm}^2e^{-m_{0^{\pm}}^2\sigma}
\nonumber  \\
&&=\frac3{8\pi^2}\int_{m_c^2}^{s_{0^\pm}}ds\,
e^{-s\sigma}s\left(1-\frac{m_c^2}{s}\right)^2\nonumber\\
&&\times
\left(1\mp\frac{2m_cm_s}{s-m_c^2}+\frac{4}{3}\frac{\alpha_s(s)}{\pi}R_0(m_c^2/s)\right)
\nonumber\\
&&+e^{-m_c^2\sigma}\left[\pm m_c\langle
\bar{s}s\rangle_0+\frac{1}{2}(1+m_c^2\sigma)m_s\langle
\bar{s}s\rangle_0\right.
\nonumber\\
&&+\frac{1}{12}\left(\frac{3}{2}-m_c^2\sigma\right)\left\langle\frac{\alpha_s}{\pi}G^2\right\rangle_0
\nonumber\\
&&+\left(\pm\frac{\sigma}{2}\left(1-\frac{m_c^2\sigma}{2}\right)m_c
-\frac{m_c^4\sigma^3}{12}m_s\right)\langle\bar{s}g\sigma\cdot
Gs\rangle_0
\nonumber\\
&&\left.-\frac{16\pi\sigma}{27}\left(1+\frac{m_c^2\sigma}{2}-\frac{m_c^4\sigma^2}{12}\right)
\alpha_s\langle\bar{s}s\rangle_0^2\right]
\nonumber\\
&&-e^{s_{0^\pm}\sigma}\left[\pm m_c\langle
\bar{s}s\rangle_0+\frac{m_s\langle
\bar{s}s\rangle_0}{2}+\frac1{8}\left\langle\frac{\alpha_s}{\pi}G^2\right\rangle_0\right]
\nonumber  \\
&&=M(\sigma), \label{eq:borelSR0}
\end{eqnarray}
for the scalar $0^+$ and pseudoscalar $0^-$ channels respectively,
where \textit{g} denotes strong coupling constant,
$G^2=G_{\mu\nu}G^{\mu\nu}$ with $G_{\mu\nu}$ being the gluon field
and $\sigma\cdot G=\sigma_{\mu\nu}G^{\mu\nu}$, and $\langle
O\rangle_0$ represents the vacuum expectation value of a local
composite operator $O(0)$ at the origin. Here, $m_s$-correction is
maintained at the first order \cite{Hayashigaki:2004hj}, and the
$\alpha_s$ correction to the perturbative contribution is given as
the functions, $R_0(x)$, as
follows\cite{Reinders:1980ik,Narison:2004hm}
\begin{eqnarray}
R_0(x)&=&\frac{9}{4}+2Li_2(x)+\ln
x\ln(1-x)-\frac{3}{2}\ln\frac{1-x}{x}
\nonumber\\
&-&\ln(1-x)+x\ln\frac{1-x}{x}-\frac{1-x}{x}\ln x, \label{eq:correc0}
\end{eqnarray}
with the Spence function $Li_2(x)=-\int_0^x dt{t^{-1}}{\ln(1-t)}$.
The running coupling constant $\alpha_s(s)$ appearing in the
perturbative terms of Eq.(\ref{eq:borelSR0}) is approximated by a
one-loop form \cite{Hayashigaki:2004hj},
$\alpha_s(s)=4\pi/[9\ln(s/\Lambda^2_{QCD})]$ with $\Lambda^2_{QCD}=(
\textrm{0.25 GeV})^2$, which is determined to reproduce
$\alpha_s(\textrm{1 GeV})\simeq0.5$ \cite{The:2000hx}. We note here
that the contributions above the threshold $s_0$ on both sides of
Eq.(\ref{eq:correc0}) are assumed to be equal to each other due to
the quark-hadron duality at high scale, and have been removed out.
In fact, our expression of Eq.(\ref{eq:correc0}) is different from
Ref.[11], namely, some condensate contributions which remain finite
above $s_0$ is thrown off besides the perturbative contribution (s.
the last line in Eq.(\ref{eq:correc0})).

In order to derive the GSRs, we make the Laplacian transform on
Eq.(\ref{eq:borelSR0}) with the formula\cite{Bertlmann:1985hk}
\begin{eqnarray}
&&\frac1{2\tau}\boldsymbol{\hat{L}}\left[\frac1{\sigma}\,e^{-(s+\hat{s})\sigma}M(\sigma)\right]
 \nonumber\\
&&=\frac1{\sqrt{4\pi\tau}}\int_0^{\infty}ds
e^{-\frac{(s+\hat{s})^2}{4\tau}}\frac{1}{\pi}\textrm{Im}\Pi(s)=G(-\hat{s},\tau).
\label{eq:gaussTr}
\end{eqnarray}
This gives a procedure to construct $G(-\hat{s},\tau)$ from
Eq(\ref{eq:borelSR0}), and hence  $G(\hat{s},\tau)$ by analytic
continuation. The resultant expressions of GRSs for the $D_s(0^\pm)$
meson currents are
\begin{eqnarray}
&&f^2_{0^\pm}m^2_{0^\pm}\exp\left[-\frac{(m^2_{0^\pm}-\hat{s})^2}{4\tau}\right]
\nonumber\\
&&=\frac{3}{8\pi^2}\int^{s_{0^\pm}}_{m^2_c}ds\exp\left[-\frac{(s-\hat{s})^2}{4\tau}\right]
s\left(1-\frac{m^2_c}{s}\right)^2
\nonumber\\
&&\times\left[1\mp\frac{2m_cm_s}{s-m^2_c}+\frac{4}{3}\frac{\alpha_s(s)}{\pi}R_0(m^2_c/s)\right]
\nonumber\\
&&+\exp\left[-\frac{(m^2_c-\hat{s})^2}{4\tau}\right]\cdot \Bigl[\pm
m_c\langle\bar{s}s\rangle_0
\nonumber\\
&&+\frac{1}{2}\left(1+\frac{2(m^2_c-\hat{s})m^2_c}{4\tau}\right)m_s\langle\bar{s}s\rangle_0
\nonumber\\
&&+\frac{1}{12}\left(\frac{3}{2}-\frac{2(m^2_c-\hat{s})m^2_c}{4\tau}\right)
\left\langle\frac{\alpha_s}{\pi}G^2\right\rangle_0
\nonumber\\
&&\pm\frac{1}{2}\left(\frac{3m^2_c-2\hat{s}}{4\tau}-\frac{2(m^2_c-\hat{s})^2m^2_c}{(4\tau)^2}\right)
m_c\langle\bar{s}g\sigma\cdot Gs\rangle_0
\nonumber\\
&&+\left(\frac{(m^2_c-\hat{s})m^4_c}{(4\tau)^2}-\frac{2}{3}\frac{(m^2_c-\hat{s})^3m^4_c}{(4\tau)^3}\right)
m_s\langle\bar{s}g\sigma\cdot Gs\rangle_0
\nonumber\\
&&-\frac{16\pi}{27}\left(\frac{m^2_c-2\hat{s}}{4\tau}+\frac{(m^2_c-\hat{s})m^2_c(3m^2_c-2\hat{s})}
{(4\tau)^2}\right.
\nonumber\\
&&\left.\left.-\frac{2}{3}\frac{(m^2_c-\hat{s})^3m^4_c}{(4\tau)^3}\right)\alpha_s\langle\bar{s}s\rangle^2_0\right]
-\exp\left[-\frac{(s_{0^\pm}-\hat{s})^2}{4\tau}\right]
\nonumber\\
&&\times\left[\pm m_c\langle
\bar{s}s\rangle_0+\frac{m_s\langle\bar{s}s\rangle_0}{2}
+\frac1{8}\left\langle\frac{\alpha_s}{\pi}G^2\right\rangle_0\right]
\nonumber \\
&&=G(\hat{s},\tau), \label{eq:BGauss0}
\end{eqnarray}
in the scalar $0^+$ and pseudoscalar $0^-$ channels, respectively.

Taking the derivative with respect to $\hat{s}$ for both sides of
Eq.(\ref{eq:BGauss0}), we get
\begin{equation}
f^2_{0^\pm}m^2_{0^\pm}(m^2_{0^\pm}-\hat{s})
e^{-\frac{(m^2_{0^\pm}-\hat{s})^2}{4\tau}}=2\tau\frac{\partial
G(\hat{s},\tau)}{\partial\hat{s}}.\label{eq:deri}
\end{equation}
From Eqs.(\ref{eq:BGauss0}) and (\ref{eq:deri}), we obtain the sum
rules for the masses of lowest resonances and couplings to the
corresponding currents
\begin{align}
m^2_{0^\pm}&=\hat{s}+\frac{2\tau}{G(\hat{s},\tau)}\frac{\partial
G(\hat{s},\tau)}{\partial\hat{s}},
\label{eq:mD}\\
f^2_{0^\pm}(s;\hat{s},\tau)&=\frac{G(\hat{s},\tau)}{m^2_{0^\pm}}
\exp\left[\frac{(m^2_{0^\pm}-\hat{s})^2}{4\tau}\right].\label{eq:fD}
\end{align}

In order to extract the values of physical quantities from the GSRs,
we use the following standard values for QCD parameters appearing in
the OPE\cite{Hayashigaki:2004hj}(see Tab.\ref{Tab:param}).
%

\begin{table}[!htp]
\caption{\small QCD input parameters used in the analysis.}
\begin{center}
\begin{tabular}{ll}\hline
Parameters & References\\ \hline
$m_s=0.11\pm0.01 \,\rm{GeV}$ & \cite{Reinders:1985im}\\
$m_c=\textrm{1.46
GeV}$ & \cite{Narison:2001mh}\\
$\langle\bar{n}n\rangle_0=(-0.225\pm0.025\,
\rm{GeV})^3$ & \cite{Reinders:1985im}\\
$\langle\bar{s}s\rangle_0=(0.8\pm0.1)\times\langle\bar{n}n\rangle_0\textrm{
GeV}^3$ & \cite{Reinders:1985im}\\
$\langle\frac{\alpha_s}{\pi}G^2\rangle_0=(\textrm{0.33 GeV})^2$ &
\cite{Narison:2004hm} \\ $\langle\bar{n}g\sigma \cdot
Gn\rangle_0=M^2_0\langle\bar{n}n\rangle_0$ & \cite{Jin:1981ig}\\ $
M^2_0=2\times(0.4\pm0.1)\, \rm{GeV}^2$ & \cite{Jin:1981ig}\\
$\langle\bar{s}g\sigma \cdot
Gs\rangle_0=M^2_0\langle\bar{s}s\rangle_0$ &  \\
$\alpha_s\langle\bar{n}n\rangle^2_0=0.162\times10^{-3}\textrm{
GeV}^6$ &  \\
$\alpha_s\langle\bar{s}s\rangle^2_0=(0.8\pm0.1)^2\times\alpha_s\langle\bar{n}n\rangle^2_0\textrm{
GeV}^6 $ & \\\hline \label{Tab:param}
\end{tabular}
\end{center}
\end{table}

For numerical calculation, we must determine the values of $\hat{s}$
and the thresholds $s_0$. To investigate the properties of the
considered resonance, the the value of $\hat{s}$ should
approximately be set to be the corresponding mass squared,
$m_{D_s}^2$, of the resonance. To suppress the continuum
contribution, we require $\hat{s}\leq m_{D_s}^2$. The conditions for
determine the value of $s_0$ are: first, it should be grater than
$m_{D_s}^2$; second, it should guarantee that there exists a sum
rule window for our sum rules. We note that the upper limit of the
sum rule window is determined by requiring the contribution of
continuum to be lower than $30\%$ of the total, while the lower
limit of that window is obtained by requiring the non-perturbative
contributions, proportional to a positive powers of $\sigma$, to be
less that $30\%$ (in fact, less than $10\%$) of the perturbative
one. Therefore, the value of $s_0$, the upper and lower limits of
the corresponding sum rule window are determined in self-consistent
manipulation.

\begin{figure}[!h]
\begin{center}
\includegraphics[angle=0,width=8cm]{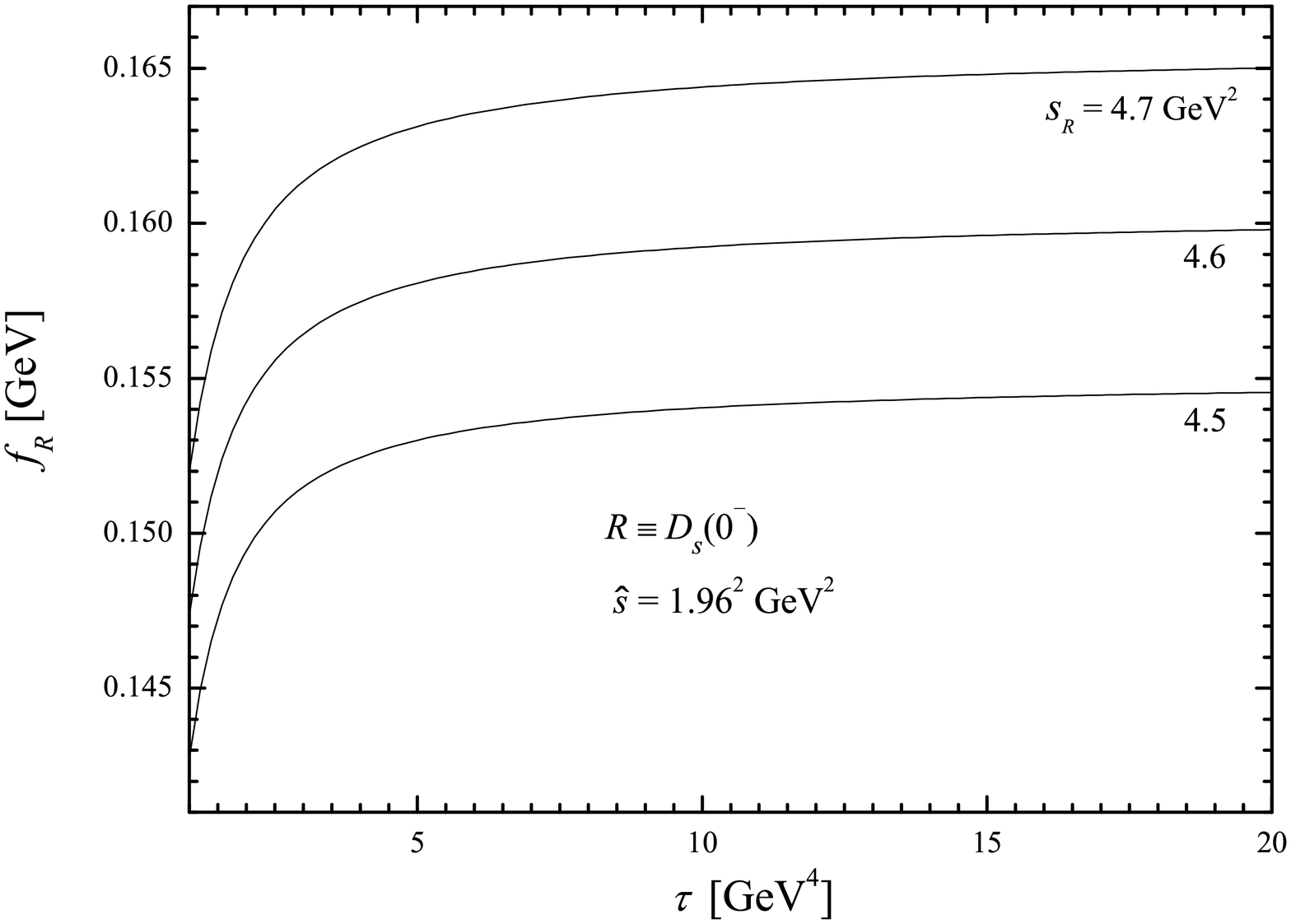}
\caption{\small The Gauss curves of $f_R\equiv f_{D_s{(0^-)}}$ vs.
$\tau$, where $\hat{s}=1.96^2\textrm{ GeV}^2$,
$s_{D_s{(0^-)}}=4.5\sim 4.7\textrm{ GeV}^2$. }\label{fig:gfDs0-}
\end{center}
\end{figure}

\begin{figure}[!h]
\begin{center}
\includegraphics[angle=0,width=8cm]{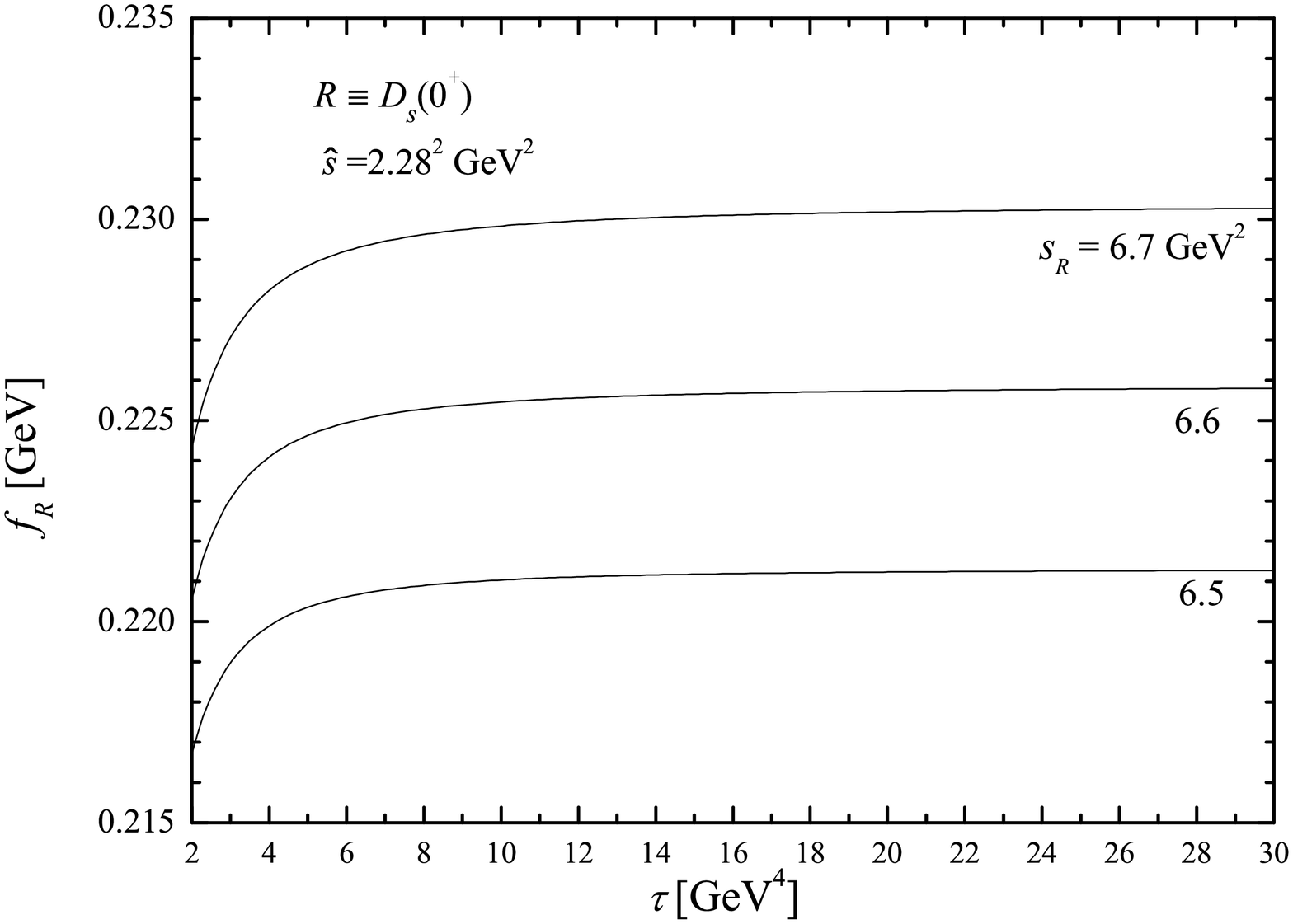}
\caption{\small The Gauss curves of $f_R\equiv f_{D_s{(0^+)}}$ vs.
$\tau$, where $\hat{s}=2.28^2\textrm{ GeV}^2$,
$s_{D_s{(0^+)}}=6.5\sim 6.7\textrm{ GeV}^2$. }\label{fig:gfDs0+}
\end{center}
\end{figure}

Then, using the GSRs of the mass and coupling constant of
$D_s(0^{\pm})$, we can get the figures and numerical results shown
below. Figs. \ref{fig:gmDs0-} and \ref{fig:gmDs0+} display the
dependencies of the calculated masses $m_{D_s(0^{\pm})}$ on $\tau$ from
GSRs, Figs. \ref{fig:gfDs0-} and \ref{fig:gfDs0+} plot the coupling
$f_{D_s(0^{\pm})}$ vs $\tau$. In Figs. \ref{fig:gmDs0-} and
\ref{fig:gmDs0+}, the thick horizontal-lines on the curves are the
Gauss stabilities of the sum rule windows. The length of the lines
corresponds to the sizes of stable regions.

\begin{figure}[!h]
\begin{center}
\includegraphics[angle=0,width=8cm]{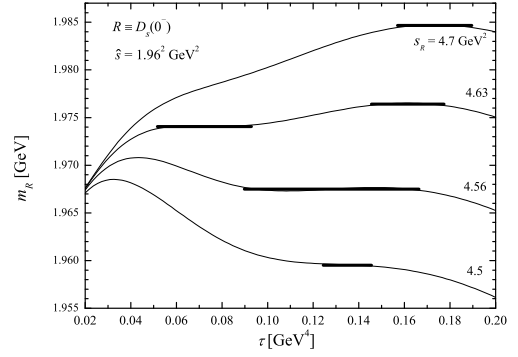}
\caption{\small The curves of $m_R\equiv m_{D_s(0^-)}$ vs. $\tau$
from GSRs, where $\hat{s}=1.96^2\textrm{ GeV}^2$,
$s_{D_s(0^-)}=4.5\sim 4.7\textrm{ GeV}^2$.}\label{fig:gmDs0-}
\end{center}
\end{figure}

\begin{figure}[!h]
\begin{center}
\includegraphics[angle=0,width=8cm]{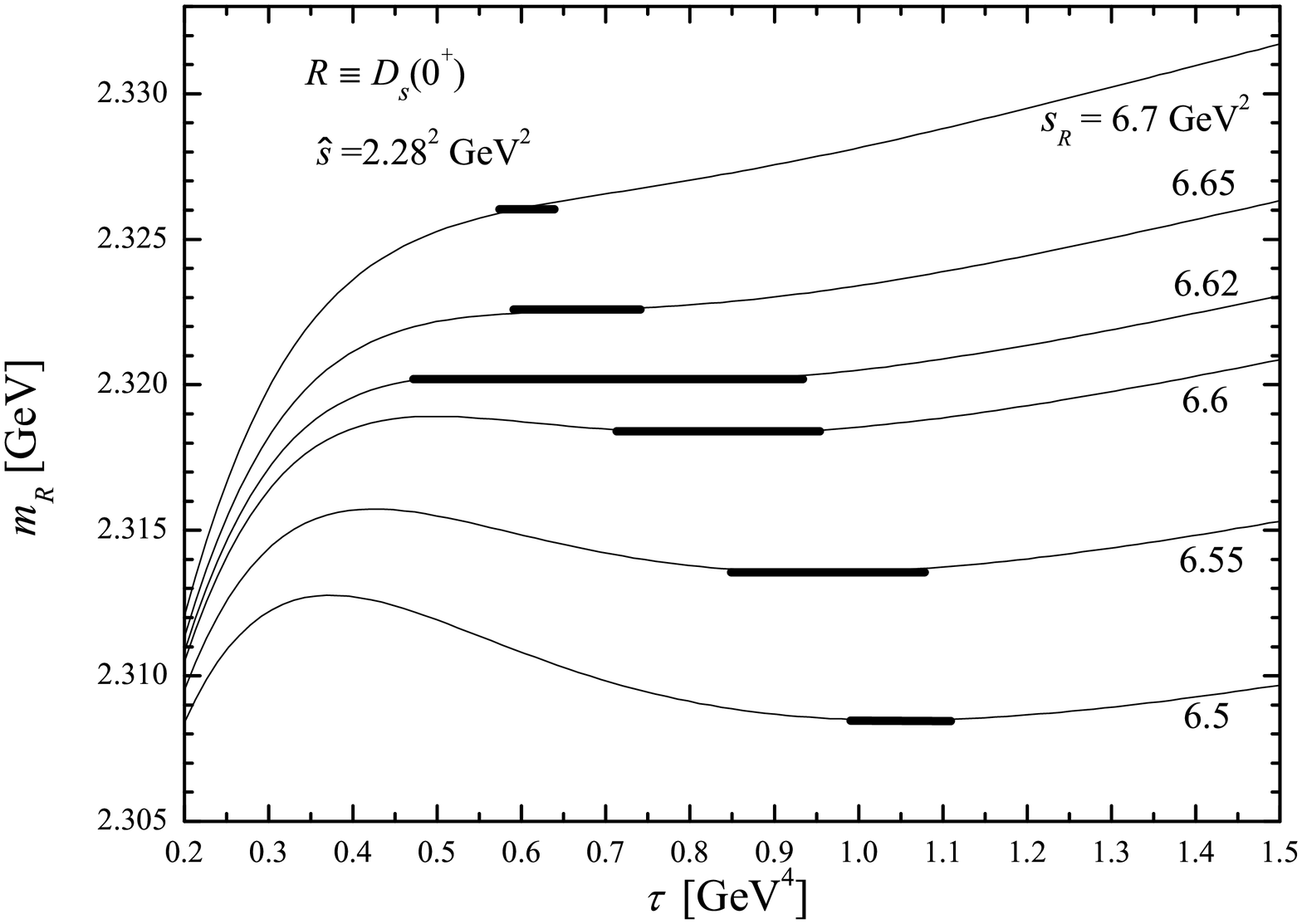}
\caption{\small The Gauss curves of $m_R\equiv m_{D_s(0^+)}$ vs.
$\tau$, where $\hat{s}=2.28^2\textrm{ GeV}^2$, $s_{D_s(0^+)}=6.5\sim
6.7 \textrm{ GeV}^2$.}\label{fig:gmDs0+}
\end{center}
\end{figure}

In $D_s(0^-)$ channel with $\hat{s}=1.96^2\textrm{ GeV}^2$, we
cannot find any Gaussian stability below $s_{0^-}=4.5\textrm{
GeV}^2$ and above $s_{0^-}=4.7 \textrm{ GeV}^2$, while between these
two values the curves are stable. From these plateau regions, we get
the resonance mass, $m_{D_s(0^-)}$, to be $1.968\pm0.016\pm0.003$
GeV, in which, $\pm0.016$ is the error from theory (here, we mean
that the upper and lower values of $m_{D_s(0^-)}$ are determined by
the upper and lower limits of $s_{o^-}$, as done in Ref.\cite{Hayashigaki:2004hj}) and
$\pm0.003$ is the error from input parameters. We can see that the
result of GSR is larger than the value of
BSR\cite{Hayashigaki:2004hj}, $1.94\pm0.03$ GeV, and closer to the
experimental value $1.969$ GeV (s. Table \ref{tab:NumReDs}). The
values of $f_{D_s(0^-)}$, $m_{D_s(0^+)}$ and $f_{D_s(0^+)}$ are
obtained in a similar way. The Numerical results are listed in Table
\ref{tab:NumReDs}. We can see from Table \ref{tab:NumReDs} cleanly
that the results of GSRs are in accordance well with experiment than
others.

For checking the self-consistency of GSRs, we also compare the
l.h.s. of Eq(\ref{eq:BGauss0}) with the r.h.s., using the center
values determined from GSRs. We have found that the two sides of
GSRs, Eq. (\ref{eq:BGauss0}), are compatible from each other very
well.

%
\begin{table}[!h]
\tabcolsep 1.5mm \caption{\small The numerical results of QCD sum
rules. For comparison, we attach experimental  avarage values
observed \cite{Vaandering:2004he, Aubert:2003ha,
Besson:2003hx,Krokovny:2003mh, Aubert:2004hb, Abe:2004hk} and the
masses of the first radial excitations predicted in
Ref.\cite{Pierro:2001hm}. (GeV)}
\begin{center}
\begin{footnotesize}
\begin{tabular}{@{}lcccc@{}}
\hline \addlinespace[1mm]
$m_R$ & (GSR)  & (BSR)\cite{Hayashigaki:2004hj}&(exp.) &(model)\cite{Pierro:2001hm}   \\
\addlinespace[1mm] \hline \addlinespace[1mm]
$0^-$ & $1.968\pm0.016\pm0.003$ & $1.94\pm0.03$ & $1.969$ & $2.700$ \\
$0^+$ & $2.320\pm0.014\pm0.003$ & $2.48\pm0.03$ & $2.317$ & $3.067$ \\
\hline\addlinespace[1mm]
$f_R$   & $f_R$ (GSR)    \\
\hline\addlinespace[1mm]
$0^-$ & $0.158\pm0.006\pm0.003$  \\
$0^+$ & $0.225\pm0.005\pm0.003$  \\
\hline \addlinespace[2mm] \label{tab:NumReDs} %
\vspace{-0.5cm}
\end{tabular}
\end{footnotesize}
\end{center}
\end{table}
%

As a summary, in the framework of the Gaussian sum rules, the masses
of the $D_s(0^\pm)$ mesons are investigated from a view-point of
ordinary light-heavy system. The GSRs for the masses of
$D_s(0^{\pm})$ mesons and the couplings of $D_s(0^{\pm})$ mesons to
the corresponding currents are derived by means of the Laplacian
transformation to the usual Borel sum rules. Using the standard
input of QCD non-perturbative parameters, the corresponding mass
spectra and couplings of the currents to the $D_s(0^\pm)$ mesons are
obtained. By comparing both sides of the GSRs, we have shown that
there exists a stability regions within which both sides of the sum
rules are matched very well. Our results are
$m_{D_s(0^-)}=1.968\pm0.016\pm0.003$ GeV and
$m_{D_s(0^+)}=2.320\pm0.014\pm0.003$ GeV, which are in accordance
well with the experimental data, 1.969 GeV and 2.317 GeV. Finally,
it worth noting that the $D^{\star \pm}_{s}(2317)$ is treated as the
lowest resonance of the $0^+$ $c{\bar s}$ meson in our calculation,
and furthermore, its mass is even lower than that of the $0^+$
$c{\bar d}$ meson estimated in the similar way (this result will be
published elsewhere). Therefore, $D^{\star \pm}_{s}(2317)$ may be
considered to be the lowest $0^+$ charmed meson in our treatment.

 \vspace{0.2cm}

\textbf{Acknowledgements}

\vspace{0.2cm}

 This work is supported by the National Natural Science Foundation of
China under Grant No. 10775105, BEPC National Laboratory Project
R\&D and BES Collaboration Research Foundation.


\begin{thebibliography}{99}

\bibitem{Drutskoy:2008} Belle Collaboration A. Drutskoy \etal~ 2008 \emph{Phys. Rev. Lett.} {\bf100} 092001


\bibitem{Drutskoy:2005} Belle Collaboration Drutskoy A \etal~ 2005 \emph{Phys. Rev. Lett.} {\bf 94} 061802


\bibitem{Mikami:2004} Belle Collaboration Mikami Y \etal~ 2004 \emph{Phys. Rev. Lett.} {\bf 92} 012002

\bibitem{Aubert:2004hb} BABAR Collaboration Aubert B \etal~ 2004 \emph{Phys. Rev. } {\bf D69} 031101

\bibitem{Abe:2004hk}BELLE Collaboration Abe K \etal~ 2004 \emph{Phys. Rev.} {\bf D69} 112002


\bibitem{Vaandering:2004he}FOCUS Collaboration Vaandering E W \etal~ 2004 \emph{arXiv:hep-ex}/0406044

\bibitem{Krokovny:2003mh}BELLE Collaboration Krokovny P \etal~
2003 \emph{Phys. Rev. Lett.} {\bf 91} 262002

\bibitem{Aubert:2003ha} BABAR Collaboration Aubert B \etal~ 2003 \emph{Phys. Rev. Lett.} {\bf 90} 242001

\bibitem{Besson:2003hx}CLEO Collaboration Besson D \etal~ 2003 \emph{arXiv:hep-ex}/0305100

\bibitem{Eidelman:2004mh}Particle Data Group Collaboration Eidelman S \etal~ 2004 \emph{Phys. Lett.} {\bf B592} 1

\bibitem{Hayashigaki:2004hj} Hayashigaki A and Terasaki K 2004 \emph{arXiv:hep-ph}/0411285


\bibitem{Pierro:2001hm} Di Pierro M  and Eichten E 2001 \emph{Phys. Rev.} {\bf D64} 114114

\bibitem{Narison:2001mh} Narison S 2001 \emph{Phys. Lett.} {\bf B520} 115

\bibitem{Narison:2005prb} Narison S 2005 \emph{Phys. Lett.} {\bf B605} 319

\bibitem{YBDai:2003} Y. -B. Dai, C. -S. Huang, C. Liu and S. -L. Zhu,  2003 \emph{Phys. Rev}. {\bf D68}, 114011

\bibitem{Reinders:1985im} Reinders  L J, Rubinstein H R  and Yazaki S 1985 \emph{Phys. Rep.} \textbf{127} 1

\bibitem{Shifman:1979ik} Shifman M A, Vainstein A I and Zakharov AV I 1979 \emph{Nucl.Phys.} \textbf{B147} 385

\bibitem{Narison:2004hm} Narison S 2004 ``\emph{QCD as a Theory of Hadrons, From Partons to Confinement}",
Cambridge University press and references therein.

\bibitem{Reinders:1980ik} Reinders L J, Rubinstein H R and Yazaki S 1980 \emph{Phys. Lett.}
\textbf{97B} 257


\bibitem{The:2000hx} The value of $\alpha_s$ at any scale can be
obtained from http://www-theory.lbl.gov~ianh/alpha/alpha.html.

 \bibitem{Bertlmann:1985hk} Bertlmann B A, Lanner G, de Rafael E 1985 \emph{Nucl.Phys.} \textbf{B250} 61

\bibitem{Jin:1981ig} Jin X, Cohen T D, Furnstahl R J and Griegel D K 1993 \emph{Phys. Rev.} \textbf{C47} 2882



\end{thebibliography}
\end{document}